# Elements of the System Theory of Time

M. Burgin

Department of Mathematics
University of California, Los Angeles
405 Hilgard Ave.
Los Angeles, CA 90095
mburgin@math.ucla.edu

**Abstract:** In the paper, elements of the system theory of time are presented, mathematical models for time are constructed, and various properties are deduced from the main principles of the system theory of time. This theory is a far-reaching development of the special relativity theory. One of the main principles of the special relativity theory is that two physical systems that are moving relative to each other have different times and it is necessary to use a correspondence between clocks in these systems to coordinate their times. Such correspondence is established by means of electromagnetic signals. In accordance with this principle, it is postulated in the system theory of time that each system has its own time. In some cases, two systems have the same time. In other cases, times of systems are coordinated or correlated. However, there are systems in which times are independent from one another.

**Key words:** time, system, inner time, external time, scale of time,

## 1. INTRODUCTION

Functioning and development of any system has temporary aspects. That is why, as the Nobel prize winner Ilya Prigogine writes (1980), time is one of the main concepts of the whole Western civilization. As it is stressed by Fields, Wright, and Harrison (1996), time dependencies and temporal constraints are an important aspect of action, and failure to meet them leads to an important class of human errors; many of the errors associated with safety critical systems have a significant temporal component. Consequently, it is necessary to have an adequate understanding of time. The conventional approach, which is prevalent now, assumes that only physical time is real and objective (Levich, 1988). However, there are different kinds of time besides the physical one: *social* (Artemov, 1987) *geological*

(Parks and Thrift, 1980), *biological* (Sutton and Ratey, 1998; Winfree, 1986), *physiological* (Winfree, 1986), *psychological* (Priestley, 1968; Golovaha and Kronik, 1984), *philosophical* (Podoroga, 1993), *astronomical* (*Le Temps,* 1968) time etc. (Zwart, 1976; Burgin, 1997). As writes Norbert Wiener (1961), one of the most famous philosophers of the 20th century Bergson lays special emphasis on the distinction between the reversible time of physics, in which nothing new happens, and the irreversible time of evolution and biology, in which there is always something new.

The problem of different time scales is urgent for the human-computer interaction. Development of interaction tools and research in this area made clear that this is a critical factor for improving quality and usability of computers (Hartson and Gary, 1992). A dynamic model of multi-system interaction is developed in (Burgin, Karplus, and Liu, 2001) aiming at the problems of time in the human-computer interaction design.

However, even in physics two kinds of time are considered (Prigogine, 1980). In philosophy, it is possible to speak about various times. Any thought or idea (in philosophy or in art) always exists in two kinds of time (Podoroga, 1993). The first kind of time in philosophy is a reign of a well to a system. But there is another 'individual' time which is 'living' inside the first one. The second time has its own order and duration, which are not reducible to the analogous properties of the first time.

Various authors considered the problem of time from different points of view. However, in each case only a partial perspective has been considered. In contrast to this, the system theory of time is aimed at the development of a systemic multifaceted approach to time and at elaboration of a mathematical theory of time. The conventional approach assumes that time is a one-dimensional quantity, made up of points, where each point is associated with a value (Hartson and Gary, 1992). The points are ordered along the dimension by their values. The common concepts of later and earlier correspond to larger and smaller values of time, respectively. This view of time is compatible with the traditional psychological, thermodynamic, and cosmic views of time (Hawking, 1988). Another approach is based on temporal intervals and denies the standard view of mapping time to points on the real number line (Prigogine, 1980). Time intervals are primary objects of observation and of learning by children. The main distinctions of the system theory of time: multiplicity of time, non-linearity in a general case, and multiplicity of formal representations.

*Multiplicity of time* means that each system has its own time. Only interaction between systems makes explicit various relations between times in each of the systems in question. In addition to this, processes going on in the same system may also develop in different times.

*Non-linearity of time* means that order between time elements (points or intervals) is not necessarily linear. For example, let us consider a cartoon that relates a story that goes on in 1900 and covers a one year period of time. Then time in this cartoon will be cyclic because each display of this cartoon repeats all events in the same sequence. Another example is given by two computer games, like Civilization, which are played on two computers that have no connections. In this case, we cannot say that one event in the first game happened earlier than another event in the second game. Thus, there is no linear time for the both games. In nature, biological time scale has topology of a circle (Winfree, 1982; 1983; 1986).

*Multiplicity of formal representations* means that time may be corresponded to different formal structures. For example, time is usually represented by a real line (as in physics or biology), system of intervals on the real line (Allen, 1984), system relations (as in temporal and dynamic logics), and natural numbers (as in theory of algorithms and theoretical models of computers: Turing machines, Petri nets, RAM, inductive Turing machines, etc.).

Different kinds of time are represented, systematized and studied in the system theory of time (Burgin, 1992; 1997; Burgin, Karplus, and Liu, 2001), elements of which are presented below. It is based on the theory of named sets (triads) and general theory of properties. Other approaches to the problem of time may be analyzed from the point of view of the system theory of time.

The system theory of time is developed in an axiomatic way as a system of principles/axioms, from which other properties of time and temporal properties of systems are derived. Principles describe different kinds of time. However, it is necessary to have in mind that only some of these principles are valid for time in general. Other principles characterize time in some classes of dynamical systems, in other words, these principles are true for some systems and false for others.

There are two groups of principles of the system theory of time: ontological principles, which represent inherent properties of time, and axiological principles, which are connected to measurement of time. Here, we give only ontological principles build mathematical models based on these principles, and deduce some properties of time from these principles.

## 2. MATHEMATICAL PRELIMINARIES

To achieve completeness of our exposition, we need some concepts from the theory of named sets (triads) and theory of abstract properties as our understanding and modeling of time is based on the concept of a triad (named set) as well as on the methods and apparatus of the theory of named sets (triads).

At first the structure of a triad was explicated in a general form in mathematics and was called "*a named set*" (Burgin,1990). This concept appeared in a process of development and unification of such important mathematical fields as fuzzy set theory and theory of multisets. Thus a new structure (which encompasses fuzzy sets and multisets) was introduced. It was a named set or, as it was later called, a *fundamental triad* (Burgin, 1997a).

**Definition 1.** A *fundamental triad* has the following structure:

$$\textbf{Entity 1} \xrightarrow{\textbf{connection}} \textbf{Entity 2} \qquad (1)$$

or

$$\textbf{Essence 1} \xrightarrow{\textbf{correspondence}} \textbf{Essence 2} \qquad (2)$$

Each component of a fundamental triad plays its unique role and has a specific name: the Entity 1 (Essence 1) is called the *support*; the Entity 2 (Essence 2) is called the *reflector* (or the *set of names*); and the connection (correspondence) is called the *reflection* (or the *naming relation*) of the fundamental triad.

In mathematics theory of fundamental triads is developed as the theory of named sets. Many mathematical structures are particular cases of named sets. The most important of such structures are fuzzy sets (Zadeh, 1965) and multisets (Aigner, 1979; Knuth, 1973; Stanley, 1986) because they are both natural generalizations of the concept of a set and have many important applications. Moreover, any ordinary set is, as a matter of fact, some named set, and namely, a singlenamed set, i.e., such a named set in which all elements have the same name. Really, there are two familiar and natural ways of constructing sets: extentional and intentional. On the one hand, in the extentional approach, given a multiplicity of objects, some or all of these objects can be conceived together as forming a set. While doing so, we baptize this set by some name. The simplest name of a set has the form $X$ or $A$. Then all elements from the set $X$ have one common name "*an element from X*". This name discerns elements from $X$ and all other entities. On the other hand, intentional method is a construction of elements that form the set. Than each element in this set is named, in some sense, and its name is the procedure (algorithm, program) by which the in question is constructed.

Thus, we see that the fundamental triad **X** has the form $(X, f, I)$ where $X$ and $I$ are some essences (entities, may be sets or classes) and $f$ is a correspondence (connection) between $X$ and $I$. Then $X$ is the support, $I$ is the reflector or the set of names and $f$ is the reflection or the naming relation of the fundamental triad **X**.

It is necessary to remark that many kinds of fundamental triads are essentially indecomposable, indissoluble units. As a consequence, in such fundamental triads the objects $X, f, I$ do not exist independently outside the triad which includes them.

Another important fundamental triad is connected with the notion of time:

**Past - Present - Future**

Here, Past is the support, Present is the reflection, and Future is the reflector of this basic time triad. It is called the fundamental time triad.

There is a similar time triad:

**before - now - later**

Theory of named sets makes it possible to give an exact definition of such a widely used notion as '*property*' and to develop mathematical theory of properties, which includes logic as its subtheory (Burgin, 1990a).

Let **U** be some class (universe) of objects, and **M** - be an abstract class of partially ordered sets, i. e., such a class that with any partially ordered set contains all partially ordered sets which are isomorphic to it.

**Definition 2.** A property P of objects from some universe **U** is a named set P = (**U**, $p$, L) where $p$ is a partial function (L-predicate), which is called the evaluation function of the property P, has **U** as the domain and some partially ordered set L∈ **M** as the codomain, which is called the scale of the property P. That is, $p$ is defined in **U** and takes values in L.

We denote the scale of the property P by Sc(P) = L, the universe of the property P by Un(P) = **U**, and the evaluation function of the property P by Ev(P) = $p$.

The most popular property in logic is '*truth*' defined for logical expressions. In classical logics only one such property as '*truth*' with the scale L = {T, F } is considered. In multivalued logics, initial intervals of the naturals with the natural ordering on them are used as the scales for this properly. For modal logics, which have only one truth property that is determined for logical expressions, modalities are expressed by means of modal operators. Another possibility to express modality is to determine different modal truth properties: '*the truth*', '*the necessary truth*', and '*the possible truth*'. For modal tense logics the quantity of such truth properties is much bigger because an individual truth property is corresponded to any operator.

Let $P_1$ , ... , $P_n$ , P be some properties on the class **U** with scales $L_1$, ... , $L_n$ , and L, correspondingly. All these scales belong to M , and a mapping $\omega: L_1 \times ... \times L_n \to L$ is defined. Here X×Y denotes the direct product of sets X and Y.

**Definition 3.** A property P with a scale L is called an n-ary ω-composition of the properties $P_1$ , ... , $P_n$ , if for any $a \in$ **U,** we have P($a$) = ω ($P_1(a)$, ..., $P_n (a)$), when $P_i (a)$ are defined for $a$ and for all i= 1, ..., n; and P($a$)= * in other cases.

Here P($a$) = * means that P($a$) is undefined.

The ω-composition of the properties is denoted by P = $\overline{P_1 , ... , P_n \omega}$.

The main logical operations: conjunction &, disjunction ∨, and implication →, are examples of ω-compositions of those properties that are determined by propositions. Thus, we have the &-composition, ∨-composition, and →-composition of logical properties.

Another class of examples is provided try physical laws, which in a lot of cases describe some equalities for properties of studied systems and phenomena, while these properties are compositions of other properties. For instance, the law of Newtonian dynamics $F = ma$ shows that the property of material bodies called "*force*" is a composition of the properties "*mass*" and "*acceleration*".

**Definition 4.** A property $P = (U, p, L)$ is equivalent to a system $Z$ of properties $\{P_i = (U, p_i, L_i); i \in I\}$ if the validity of the inequality $P_i(a) \neq P_i(b)$ for some $i \in I$ implies $P(a) \neq P(b)$ for any $a, b \in U$, and vice versa.

Composition of properties makes possible to prove the following result (cf.

**Theorem 1.** *For any system* $Z = \{P_i = (U, p_i, L_i); i \in I\}$ *of properties, there is a property* $P = (U, p, L)$ *equivalent to* $Z$.

## 3. INTRODUCTION TO THE SYSTEM THEORY OF TIME

Now, after we have given necessary mathematical preliminaries, we can consider time and its properties. Here we give only ontological principles of the system theory of time, which reflect inherent temporal properties of various systems (physical, biological, social, etc.). Axiological principles, which are dealing with measurement of time, are presented in (Burgin, 1997).

An important peculiarity of a system approach to time is the emphasis that is made on the necessity to consider time not as an absolute essence but to relate time to some definite system. So, let us take some system $R$ and discuss relations between time T and $R$.

**Principle O1.** *Any system has its own time*.

**Remark 1**. However, time in two different systems may be isomorphic.

When applied to physical systems, Principle O1 implies that in different coordinate systems time is different and it is necessary to correspond one local time to the other in order to coordinate events in these systems. Special theory of relativity determines how to do this coordination. In this sense, the system theory of time is an extension of the special theory of relativity.

To derive other properties of time from Principle O1, we apply system theory. According to its principles, any system **R** is not unique, if we exclude the system of everything, which in some sense is ill-defined. Consequently, there is another system **Q**, which is not included in **R**. In general, we have the following triad

$$\mathbf{R} \xleftrightarrow{\text{interaction}} \mathbf{Q} \qquad (3)$$

Principle O1 implies that **R** has some time $T_\mathbf{R}$, **Q** has some time $T_\mathbf{Q}$, and interaction as a system has some time $T_\mathbf{RQ}$. If we consider $T_\mathbf{R}$, $T_\mathbf{Q}$, and $T_\mathbf{RQ}$ with respect to **R**, then we come to the following result.

**Theorem 2.** *Three types of time exist for an arbitrary system* **R**: *an internal, connection, and external time.*

Here, $T_\mathbf{R}$ is an internal time for **R**, $T_\mathbf{Q}$ is an external time for **R**, and $T_\mathbf{RQ}$ is a connection time for **R**.

Let us give exact definitions

**Definition 5.** Internal time $T_\mathbf{R}$ of **R** is the time, in which the whole system **R** or some of its subsystem functions.

**Definition 6.** External time $T_{e\mathbf{R}}$ for **R** is an internal time of its environment as a system.

From this definition, it follows that an external time for **R** is an internal time of any system that is not included in **R**.

**Definition 7.** Connection time $T_{c\mathbf{R}}$ for **R** is the time in which the processes of interaction of **R** with its environment go on.

An external time for a system **R** is the inner time $T_{e\mathbf{R}}$ of any system **Q**, which is not a subsystem of **R**. However, the system **Q** may be unrelated to **R**. In this case, times in these systems also look unrelated to **R**, while by Definition 6 they are attributed to **R**. Thus, to eliminate this difficulty, we introduce a proper external time $T_{pe\mathbf{R}}$.

**Definition 8.** A proper external time $T_{pe\mathbf{R}}$ for a system **R** is the inner time $T_\mathbf{Q}$ of any system **Q**, which contains **R**.

A union of two systems may be considered as a system containing both initial systems. So, we have the following result.

**Proposition 1**. *For any system **R**, there is a proper external time.*

Internal, or inner, time $T_{IR}$ is an inherent property of the system **R**. Several kinds of inner times are connected with such a system as a human being. There is biological time, which is the inner time of an individual when we consider this individual as a biological (living) system. Besides, researchers distinguish physiological time if the organism of a human being is considered on the level of biochemical (physiological) process (Winfree, 1986). If we consider a human being as a biochemical system, then such a kind as the biological time is the primary one (Sutton and Ratey, 1998; Winfree, 1986). Psychological time appears on the level of personality (Burgin, 1992), while the behavioral level of individual is related to perception time, as well as to reaction time (Bragina and Dobrohotova, 1988).

Existence of several inner times in one individual implies that the age of a person is not a unique number, which is equal to the length of physical time from the birth of this person to the given moment. This is only the chronological age, which is not the most important for a human being. Much more important are biological, physiological, and psychological ages of a person. Biological age reflects how the organism of this person is functioning. Physiological age represents how biochemical reactions are going in the organism of this person. Psychological age reflects how the psyche of this person is functioning.

A source of existence of external time $T_E$ is that the system **R** is not unique – there are other systems. Time in any of these systems is an external time $T_{eR}$ for **R**. For example, any living being **B** has its internal time. At the same time, **B** is a part of nature and lives in physical time, which is external time $T_E$ for **B**.

Connection time $T_{cR}$ is the time in which goes interaction. It exists only for such pairs of systems, which interact. For example, connection time $T_{cR}$ of a person **B** with the car, this person is driving, is neither the internal time of **B** nor the physical time, which is the external time for **B**. Time $T_C$ depends on perception and reaction of the person. Consequently, it might be different for different persons even with the same car.

Existence of three types of time is a peculiarity of both nature and society. To reflect and model inner time in physical systems, a mathematical construction was introduced in

(Prigogine, 1980). Besides, inner time is discussed and utilized in the chronogeography (Parks and Thrift, 1980), as well as in biology and physiology (Sutton and Ratey, 1998; Winfree, 1986). Moreover, experiments demonstrate relative independence of of the inner biological time from the external physical time (Winfree, 1982).

To reflect properties of different kinds of time, the system theory utilizes principles, which represent the most essential properties. We begin with principles for inner time. The main principle connects the flow of the inner time in a system with changes going on in this system.

**Principle OI 1.** *Time* $T_R$ *in a system* **R** *is a labeled/indexed set (in particular, a sequence) of changes in* **R.**

For simplicity, we denote $T_R$ by T.

Alike idea is suggested by Zwart (1976), who defines time as a sequence of all events. However, this approach has several shortcomings. First, the notion of event is not sufficiently exact for a scientific concept. Second, Zwart assumes existence of the absolute time. Relativity theory invalidated idea of the absolute time, demonstrating that even physical time in different systems might be different. According to Zwart, there is only linear time, while, for example, biological time is cyclic.

**Definition 9.** A property P of **R** is called an ontological base of time T if changes of P determine the flow of time T.

That is, any change of P is implied by a change of T and any change of T on a unit element corresponds to an elementary change of P, i.e., one unit adds to the quantity of time. An ontological base P of T is denoted by P = Ont (T) and T is called the P-time of the system R.

In hierarchic systems on different levels of hierarchy, specific properties exist, which are ontological bases for the corresponding kinds of time.

In general, it is possible to correspond some inner time T of **R** to any property P of **R** taking P as the ontological base of T. But formally, it does not mean that any inner time of **R** is of the same kind.

Existence of an ontological base for an inner time is stated in the following principle.

**Principle OI 2** (`the general ontological principle`). *For any system* **R** *and any time* $T_R$ *ontological base* Ont ($T_R$) *exists*.

From this principle follows an important consequence: a moment (a unit) of inner time does not mean a position, in which the system **R** is in some state, but a point (a unit) at which a change of **R** is going on. Only for an external time, we may say that **R** is in a given state at some moment.

Nevertheless, a moment (a point, an interval) $t$ of time $T_\mathbf{R}$ may be corresponded to two states of **R** : the source state $u$ and the final state $v$. As a result, the model of I has the form of the named set $t = (u, ch, v)$ where ch is the change of **R** which transforms $u$ into $v$ at the moment $t$. Here $u = S(t)$ is the support of $t$ and $v = N(t)$ is the reflector of $t$. If t is a moment (a point) of time $T_\mathbf{R}$, then we denote it by $t \in T_\mathbf{R}$. So, the scale $Sc(T_\mathbf{R})$ of the property "time," denoted by the letter T, is a system of named sets, and each interval of the system $Sc(T_\mathbf{R})$ is represented by some named set.

**Proposition 2.** *Representation of intervals by named sets is one-to-one correspondence if and only if the* $Sc(T)$ *is a linearly ordered set*.

For a physicist (cf., for example, (Einstejn, Lorentz, Weil, and Minkowski, 1923)), time is something that is measured by clocks. Such understanding is implied by the following principle.

**Principle OI 2c** (the constructive ontological principle). $T_\mathbf{R}$ *is a set (a sequence) of detectable states of some subsystem* **C** *of* **R**, *which is called the clock for* $T_\mathbf{R}$.

Studying and discussing physiological and biological times scientists introduced such term as "inner clocks" of an organism (Richter, 1968; Winfree, 1986). They relate these clocks to circadian rhythms in the brain (Winfree, 1983; Takahashiand Zatz, 1982).

More restricted is the following principle.

**Principle OI 2d** (the strict ontological principle). *Any inner time* $T_\mathbf{R}$ *in a system* **R** *is the system of changes of some property* P *of* **R**.

From this principle (but not from Principle 0I 1), the following property of time follows.

**Principle OI 3** (the reality principle). *Time* $T_\mathbf{R}$ *in a system* **R** *is a real (binary) property of* **R**, *i.e.*, $T_\mathbf{R} = (\mathbf{U}^2, it, L_T)$.

There are different scales of time and of models of time. The most popular scale is the linear model of time T, in which the scale $L_T$ is the space **R** of all real numbers or its finite or infinite interval. However, in ancient societies a cyclic model of time was prevalent in ancient societies (Kosareva, 1988). As it is written in Ecclesiastes, "That which hath been is that which shall be."

There has been also a debate whether the scale $L_T$ of time is continuous or discrete. Although continuous time scale has been always more popular, esoteric medieval directions, as well as some classical philosophical schools in India, preferred discrete conception of time (Kosareva, 1988). Now some theories of quantum physics also utilize discrete models of time (Blokhintsev, 1982; Vyaltsev, 1965).

It is interesting to mention that Eddington suggested a hypothesis (cf. Chernin, 1987) that time is one-dimensional only in some cosmic neighborhood of the Earth. Time may be two-dimensional in some very distant domains of the universe.

Any change of a system **R** is a change of at least one of its properties. So, there is a system **Z** = ( $P_i$; i∈ I) of properties $P_i$ such that any change of **R** is a change of some $P_i$ from **Z**. By the theorem 1 there is a property P which is equivalent to the system **Z**. In particular, any change of **R** is a change of P. As a consequence, we have the following result.

**Proposition 3.** *Principle* 0I 2a (*and thus, Principle* 0I 2) *follows from Principle* OI 1.

**Principle OI 4** (`the dynamic principle`). *Time* T *changes only when the base* Ont(T) *changes*.

In science time connected with changes of the knowledge system is the most important. It is called the *gnostic* time and is determined by changes in the knowledge system **K** of science. Now there are no means of precise definition of the gnostic time because means of knowledge measurement and even of its evaluation are not developed enough. But development of such means will provide a possibility for elaboration of clocks for the gnostic time.

Taking instead of the whole knowledge system of science its subsystem related to some object field (for example, physics or science of science). We obtain the notion of the field gnostic time.

**Principle OI 5** (`the exactness principle`). *If after some changes the value of* Ont(T) *is the same as it was before these changes, then the value of time* T *is also the same as it was before.*

**Principle OI 6** (`the coexactness principle`). *If after some changes the value of time* T *is the same as it was before these changes, then the value of* Ont(T) *is also the same as it was before.*

Let us consider the external time $T_{eR}$ for a system **R**. The following principle depicts connections between internal and external time.

**Principle OE 1** (`the external ontological principle`). *There is a property* P *of* **R**, *which is reduced to* $T_{eR}$ *and is the ontological base of the internal time* $T_R$ *in* **R**.

Systems are usually represented in a parametric form. That is, a system **R** is reflected in knowledge systems (like theories) by means of a collection of parameters or attributes. Values of these parameters/attributes give the states of the system **R**. Any parametric/attributive representation is a representation by means of properties because parameters/attributes are some kinds of abstract properties. Moreover, the general theory of properties makes possible transformation of any symbolic representation of the system **R** into an attributive representation (Burgin, 1990a). In it, a collection of properties **P(R)** is corresponded to **R** and values of these properties reflect the states of the system **R.** Theorem 1 implies the following result.

**Theorem 3.** *Any parametrical/attributive representation of a system* **R** *is equivalent to a representation of* **R** *by means of a single property.*

An ontological base Ont(T) may be treated as the state vector (a vector representation) of **R** in some state space. It may be formalized in the following manner.

Let T be a time in a system **R** and $V_R$ be a phase space (a state space) for P. We fix some set $Ch \subseteq V_R \times V_R$. Elements of set *Ch* are called changes of states of **R**.

**Definition 10.** T = ( $V_R \times V_R$, *tm*, $L_T$ ) where Ont(T) = ( {**R**}, p, $W_R$ ), $W_R$ is an image of a projection *pr*: $V_R \to W_R$, and *Ch* is mapped by *pr* isomorphically.

**Remark 2.** In quantum mechanics and in statistical mechanics the set $F(V_R)$ is taken instead of $V_R$ for the parametric description of *R*. That is, a wave function is considered in quantum mechanics and a density in a phase space in statistical mechanics giving a description of, a system. In this case T = ( $F(V_R) \times F(V_R)$, tm, $L_T$ ).

If $L_T = \{1, 0\}$, then T is a representation of an operator, which may be denoted by the same letter T. When this operator satisfies the equality $-i[L, T] = 1$ where $L$ is the Liouville operator, then T is the operator of inner time defined by Prigogine (1980). Thus, the inner time of Prigogine is also included in the context of the system theory of time.

It is an interesting problem to consider such an operator of inner time for different fields of science and to compare ft with the ordinary physical time that is used for obtaining a dynamical picture of science.

Functioning of **R** is a sequence (in a general case, a partially ordered set) of changes of states of **R**. Any change of states is a transition from some state of **R** to another one.

Then Principle OI 5 states that the system **R** is in one and the same state in this space if and only if the corresponding moments of time are also equal.

**Proposition 4.** *The functioning of a system **R** results at least once in equal states (relative to all properties of **R**) if and only if any inner time T of **R** has a loop.*

Let P be a property (or, equivalently, cf. Theorem 1, a system of properties) of **R**. Then $P = (U_R, p, L)$ where $U_R$ is the collection of all states of **R**, and the scale L is a phase P-space of **R**. For example, for a material point x its kinematics space is the ordinary three-dimensional Euclidean space. An arbitrary phase space of **R** is denoted by $V_R$.

**Definition 11.** A property P of **R** is called complete (with respect to **R**) if in different states of R the corresponding values of P are also different.

**Definition 12.** A property P of **R** is called dynamically complete if any change of **R** determines a change of P.

**Proposition 5.** *Any complete property* P *is dynamically complete (with respect to the same system).*

**Remark 3.** The converse is not true. Realty, to any system R it is possible to correspond a dichotomic property $0 = (U. q, L)$ with $L = (1, 0)$, and such that 0 changes its value if it changes its state. Such a property is called a diadic fixer of changes. Any clock is functioning in a similar way, only the scale of the clock contains more than two elements. This scale consists of all positions of the hands on the face of the clock.

The considered dichotomic property 0 is dynamically complete but not complete with respect to **R** if **R** has more then two states.

Let H be a set with a binary relation $\leq$ on it.

**Definition 13.** An H-trajectory **Tr** of **R** is a functional named set **Tr** = (H, *tr*, $V_R$).

Protocols in networks are examples of H-trajectories. Thus, the system theory of time provides a theoretical background for many urgent problems of the INTERNET.

**Definition 14.** An H-trajectory **Tr** describes functioning of **R** the relation ≤ represents transitions of **R** from one state to another.

**Definition 15.** An H-trajectory **Tr** is called T-temporal if for any two different points *x* and *y* from the image of the trajectory mapping *tr* and x --y imply p(*x*) and p(*y*) are different elements from the scale of the property P = OntR (T).

**Proposition 6.** *For any linear, coherent with* T, *T-temporal* H-*trajectory* **Tr** *the time* T *is linear ordered along* **Tr**.

This reflects one of the most frequent illusions that a sequence of moments of time determines the order of events. On the contrary, in reality an order of events (of changes) induces the corresponding order on the scale of time, and thus, determines the direction of the flow of time.

Reversibility of time is an important property for HCI because if it is possible to reverse time in interaction, we can correct our mistakes and/or choose a better way of action by experimentation. Some programs have means for reversibility of time. For example to reverse time in interaction, they use such operations and commands as "*Undo*" or "↺." According to Principle OI 5, this brings the connection time $T_{cR}$ of a system **R**, which interacts with the computer, one unit back. Such a system **R** may be a user or the text processed by the computer and the program in this case is word processor.

The system theory of time gives some conditions for time reversibility.

Let us assume that Principle 0I 6 is valid.

**Proposition 7.** *Reversibility of inner time* T *implies iteration of the states of* **R** *with respect to* Ont (T).

**Remark 3.** Properties of reversibility and uniqueness of time depend on ontological axioms.

**Proposition 8.** *Uniqueness of all moments of time is equivalent to absence of loops in the structure of time.*

**Proposition 9.** *Reversibility of time* T *implies absence of uniqueness of all moments of time, i.e. some moments of time* T *coincide.*

**Definition 16.** A system **R** is called

a) primitive if it is completely represented by a single one-dimensional property;

b) multidimensional when any property, which completely represents P, is multidimensional (Burgin, 1990a).

**Proposition 10.** *In a multidimensional system **R** there are different inner times or, equivalently, time may be multidimensional, as an abstract property.*

Many examples of this situation are demonstrated in (Burgin, 1997) where a complex natural system is treated as a complex of hierarchical subsystems. Thus, in natural systems different kinds of time appear. Each kind corresponds to some hierarchical level of the given system. But, as states Proposition 10, different times may exist even on one hierarchical level. Relativity theory gives examples for such situations (Einstejn, Lorentz, Weil, and Minkowski, 1923). If we consider a physical system **R** consisting of two subsystems which are moving with a very big velocity in opposite directions.

There is a hypothesis (Winfree, 1986) that an individual has meny different inner clocks that induce different times in subsystems of the organism.

Moreover, if $P = (U, p, L)$ is a property of a system **R**, then the following result is valid.

**Theorem 4.** *It is possible to determine some time $T_P$ of **R** for which $P = Ont(T_P)$.*

<u>Proof</u> To do this, let us take some set $E$ for which a one-to-one correspondence $k: L \to E$ exists. Then $T_P = (U, t_P, E)$ where $t_P(u) = k(p(u))$. A direct test of principles 012 - 014 and definitions demonstrates that $P = Ont(T_P)$. Theorem is proved.

Any such time T, is called an inner attributive time of **R**. Existence of different kinds of time in a single system **R** makes possible to compare these kinds, and in such a way to explicate regularities of the functioning of **R**.

## CONCLUSION

An important problem of time coordination arises from existence of distinct time scales in different systems. For a system **R** to interact properly with another system **Q**, it is necessary to have a time coordination function $t_{CRQ}$. This function maps homomorphically the time $T_Q$ in the system **Q** into the time $T_R$ in the system **R**. Homomorphically means that $t_{CRQ}$ preserves some chosen structures (at least, the partial order) in the scale of $T_Q$.

The special relativity theory is based on time coordination in physical systems that are moving relative to each other (Einstejn, Lorentz, Weil, and Minkowski, 1923).

An interesting and at the same time important example of time coordination gives us people's estimation of the speed of a moving vehicle. To estimate speed, a person has to coordinate her/his inner time with time of the environment in which the vehicle moves. Such coordination is achieved by assessing changes in the environment.

Not in all cases, such assessment is correct. For example, scientists observed that when a person is driving in a fog, she or he perceives that the speed of her/his car is essentially less than the real speed. However, without fog the same people give correct estimations of the speed.

The general theory of time suggests an explanation to this phenomenon:

While driving in the fog, a person can see much less changes of the environment than in a situation when the same person is driving in a good weather. These changes define the time sensation for this person. According to the general theory of time, less number of changes implies slower movement of time in the environment. Inner time of this person stays the same and she/he projects this time on the environment. In other words, the person assumes that the same amount of changes always corresponds to the same time interval. So, when driving in the fog, it seems that the vehicle, which exists and moves in this external time, moves slower.

It is important to find properties of time coordination for different systems. It possible to find some results in this direction related to computers in (Burgin, Karplus, and Liu, 2001).

To conclude, it is necessary to emphasize that the problem of time is a scientific and not a purely logic-linguistic problem, as think some researchers in this area. Only synthesis of mathematical modeling, logical reasoning, and scientific observation and experimentation can explicate the true meaning of time and give an efficient and adequate theory of time. The correct question is not "What time is?" but "What is the structure, properties and functions of time?" Science, for example does not asks what matter is but investigates how physical reality is organized.